\documentclass{webofc}
\usepackage{graphicx}
\usepackage{subcaption}
\usepackage[varg]{txfonts}   
\usepackage{hyperref}
\usepackage{url}
\usepackage[normalem]{ulem}
\hypersetup{colorlinks=true,citecolor=blue,urlcolor=blue,linkcolor=blue}
\newcommand{\Dipper}{\textsc{McDipper}}

\begin{document}
\title{Effects of Subnucleonic Fluctuations on the Longitudinal Dynamics of Heavy-Ion Collisions}

\author{Oscar Garcia-Montero\inst{1,2} \and Sören Schlichting\inst{1} \and Jie Zhu\inst{1,3}\thanks{\email{jzhu@physik.uni-bielefeld.de}}}

\institute{Fakultät für Physik, Universität Bielefeld, D-33615 Bielefeld, Germany \and Instituto Galego de F\'isica de Altas Enerx\'ias IGFAE, Universidade de Santiago de Compostela, E-15782 Galicia, Spain \and
Institute of Particle Physics and Key Laboratory of Quark and Lepton Physics (MOE), Central China Normal University, Wuhan, 430079, China}

\abstract{%

It is well understood that subnuclear fluctuations in the initial state of heavy-ion collisions have an important impact on the creation of long-range correlations in the transverse plane. This is also true for the creation of particle correlations along the beam direction, which can be measured in particle detectors, e.g. through longitudinal decorrelation observables. In this work, we  study the emergence of long-range rapidity structures in Pb+Pb collisions using a hybrid model connecting the 3D resolved {\Dipper} initial state model to a (3+1)D viscous hydrodynamics framework CLVisc. We include different sources of fluctuations at the (sub-)nucleon level and present the effects of their inclusion on the longitudinal structure of relevant observables, focusing in this proceedings paper on charged hadron multiplicities, baryon stopping, directed flow and flow decorrelation. We find remarkable agreement to the experimental data regarding directed flow and the rapidity resolved charge particle multiplicity. }

\maketitle

\section{Introduction}

Relativistic heavy-ion collisions offer a unique environment to study strongly interacting matter under extreme conditions, where the formation of a quark–gluon plasma (QGP) can be probed through final-state observables~\cite{Jacak:2012dx}. Theoretical modeling of such collisions typically employs a multi-stage approach: (i) an initial energy and charge deposition, (ii) a pre-equilibrium phase, (iii) viscous hydrodynamic expansion of the QGP, and (iv) a late hadronic rescattering stage \cite{Elfner:2022iae}. While the later stages are now well constrained, significant uncertainties remain in the description of the initial state and its transition to local equilibrium.

While great progress has been achieved in the last decades, a large array of initial condition models assume approximate boost-invariance, which is suitable near midrapidity at high collision energies but insufficient for describing the longitudinal structure revealed in experiments, particularly when describing small systems (O+O, Ne+Ne collisions) and asymmetric systems (p+A, d+A). Observables such as baryon stopping, rapidity-dependent anisotropic flow, and flow decorrelations across pseudorapidity demonstrate the necessity of more realistic three-dimensional initial conditions~\cite{Bozek:2010vz,ALICE:2010mty}.
A particularly interesting aspect is the role of fluctuations at different scales. While nucleon-level fluctuations are known to be crucial for explaining collective phenomena and centrality dependence in heavy ions, subnucleonic fluctuations are necessary for describing fluctuation-driven observables or for smaller collision systems~\cite{PHENIX:2018hho}. However, most studies have primarily examined the effects of subnucleonic fluctuations in the transverse plane, while the impact on longitudinal decorrelation observables has received comparatively less attention.

In this short communication, we summarize results from a recent study of Pb–Pb collisions at  $\sqrt{s_{\rm NN}}= 2.76$ TeV, using the {\Dipper} model with explicit subnucleonic fluctuations coupled to the (3+1)D viscous hydrodynamic framework CLVisc. We focus on four representative observables: the rapidity dependence of charge particle yields as well as the net proton rapidity distributions, directed flow and flow decorrelation.

\section{Model Description}

As stated in the introduction, our hybrid model contains the initialization of the QGP state, which is performed on an iso-$\tau$ surface at a proper time $\tau=0.6$ fm and is subsequently evolved in (3+1)D until freeze-out using CLVisc's  Cooper-Frye prescription.\vspace{2mm}

\noindent\textbf{Rapidity-resolved initial state model: The {\Dipper}}

The {\Dipper} framework is based on the $k_T$-factorization framework of the color glass condensate\footnote{For a review of the CGC framework, we refer to Ref.~\cite{Garcia-Montero:2025hys}} (CGC). Gluon production is described via single gluon production formulas, while quark energy and charge deposition arises from multiple scattering and stopping of collinear quarks~\cite{Garcia-Montero:2023gex,Garcia-Montero:2023opu}. The latest version, {\Dipper} v1.2, 
incorporates subnucleonic fluctuations by resolving nucleons into Gaussian hotspots, whose positions fluctuate event by event. The strength of fluctuations at the (sub-)nuclear  positions can be chosen to also fluctuate by sampling the normalization of the nucleon (hotspot) strengths from a log-normal distribution centered around unity, which we call thickness fluctuations. This results in a granular 
and asymmetric initial energy deposition.

We compare the effects for five different sets of fluctuations  included into the {\Dipper} initial state\footnote{For model information and parameter choice, please see Ref. \cite{Garcia-Montero:2025bpn}}: 
 \textbf{(i)} smooth nucleons (no subnuclear fluctuations, denoted as nucleon),
\textbf{(ii)} three hotspots (denoted as hotspots),
\textbf{(iii)} thickness fluctuations at the nucleonic level (denoted as nucleon fluctuations $\sigma= 0.637$),
\textbf{(iv+ v)} three hotspots and thickness fluctuations of two different magnitudes $\sigma= 0.637, 1.2$.\vspace{-4mm}

\subsection{Hydrodynamic evolution: CLVisc}
\vspace{-2mm}

The event-by-event initial conditions are evolved with the (3+1)D viscous hydrodynamic model CLVisc~\cite{Wu:2021fjf, Pang:2018zzo}
. The code solves energy–momentum and baryon number conservation including second-order Israel–Stewart-type terms for shear stress, bulk pressure, and baryon diffusion. 
Transport coefficients are chosen following Bayesian analyses of LHC data, while the equation of state (NEOS-B) incorporates baryon chemical potential effects.

Particlization is performed at an energy density of $\epsilon_{\text{frz}}=0.266~\text{GeV/fm}^3$ 
using the Cooper–Frye prescription. Hadrons are sampled with corrections for shear and baryon diffusion, and resonance decays are included. Although a hadronic afterburner is not yet implemented, the framework provides robust qualitative predictions.\vspace{-3mm}

\begin{figure}[t]
	\centering
	\begin{subfigure}[b]{0.4\textwidth}            
		\includegraphics[width=\textwidth]{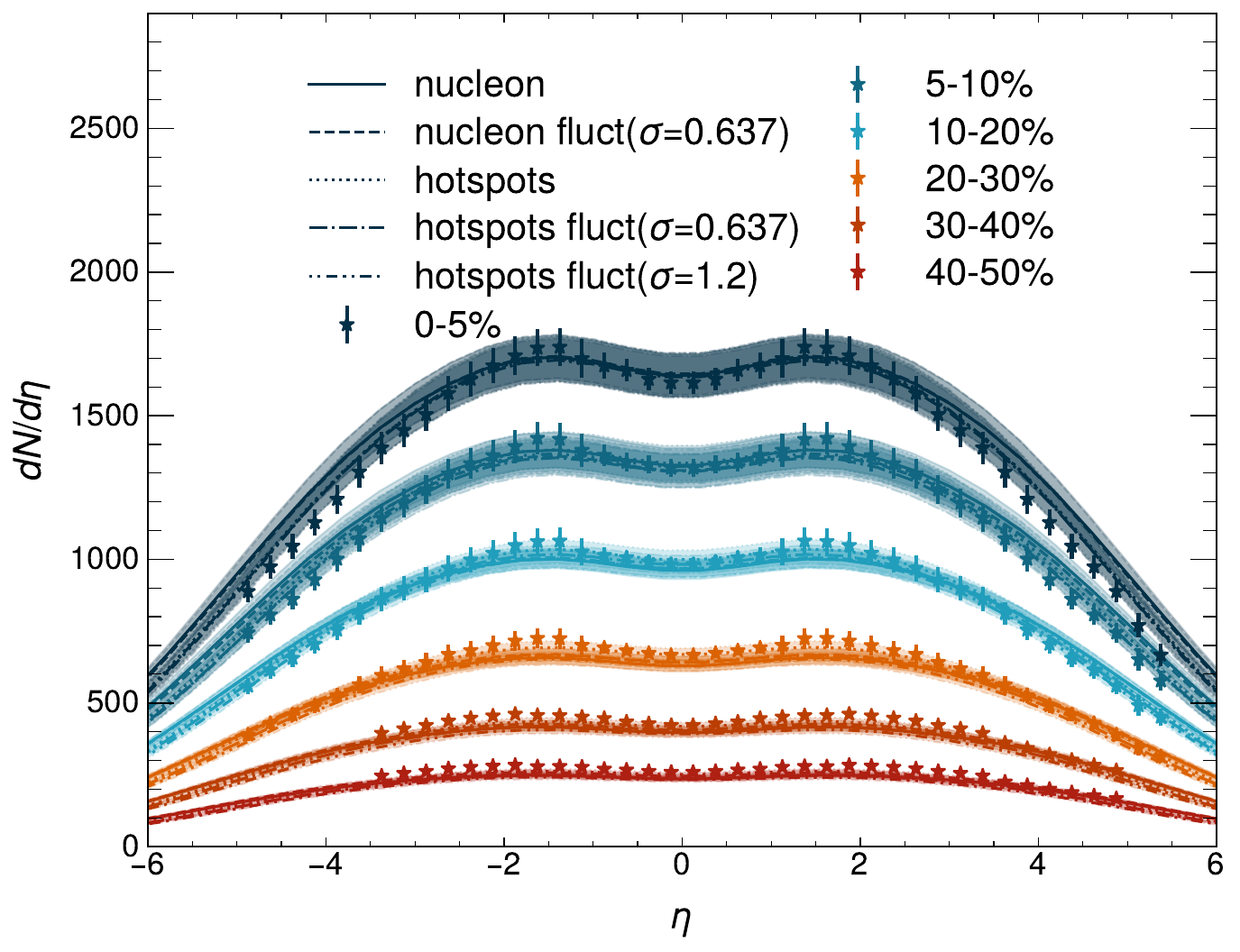}\vspace{-5mm}
		\label{fig:dnchdeta}
	\end{subfigure}%
	\begin{subfigure}[b]{0.4\textwidth}
		\centering
		\includegraphics[width=\textwidth]{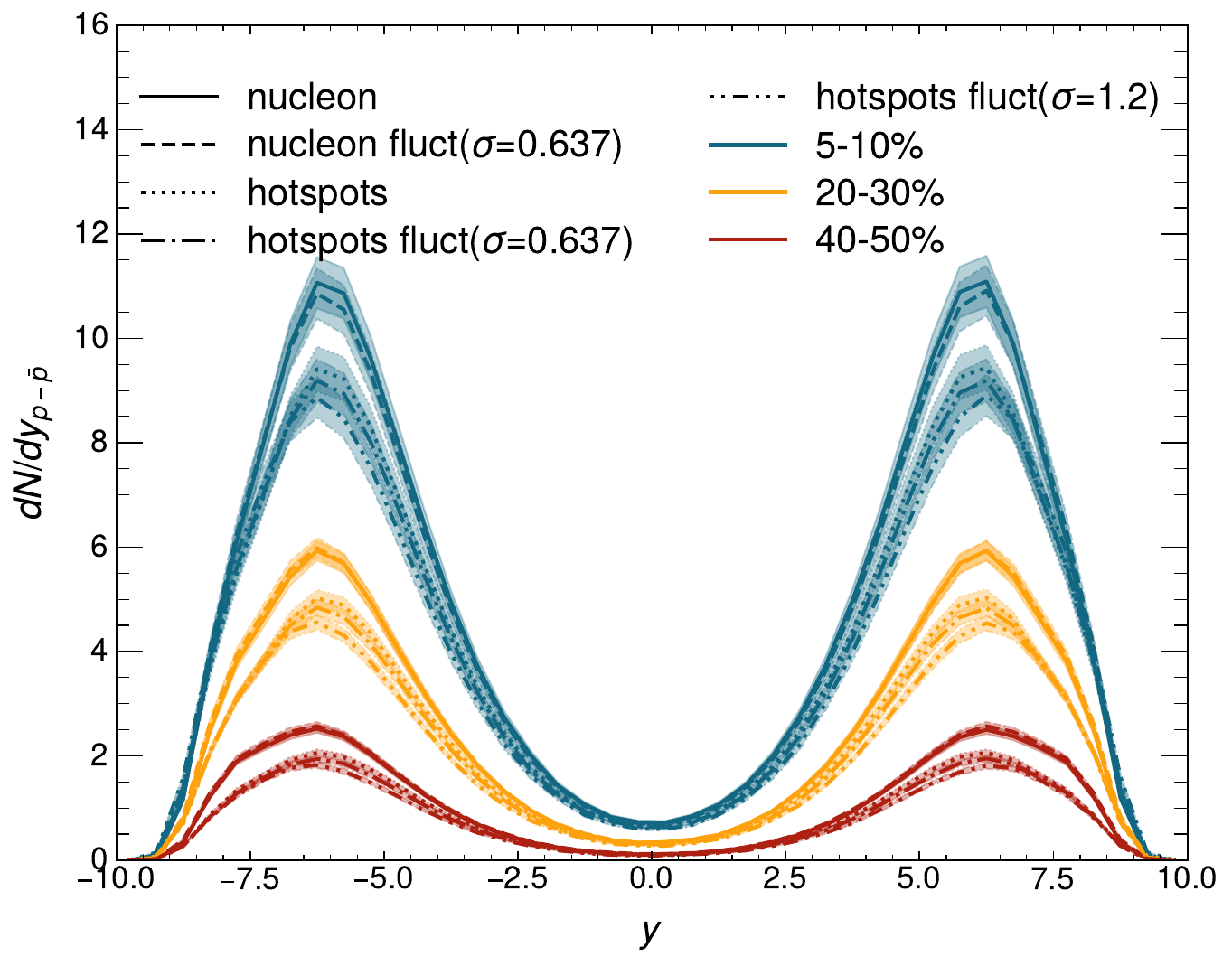}\vspace{-5mm}
		\label{fig:dndy}
	\end{subfigure}
	\caption{ (\textit{Left}) Charged hadron multiplicity distribution as a function of pseudorapidity $\eta$ for Pb-Pb collisions at 2.76~TeV compared 
		to ALICE data~\cite{ALICE:2015bpk}. (\textit{Right}) Net-proton multiplicity for same system. For readability only the 5-10,20-30 and 40-50\% centrality classes are presented.}\label{fig:mult}\vspace{-6mm}
\end{figure}

\section{Results}
\noindent \textbf{Charged hadron multiplicity}:
Fig.~\ref{fig:mult} shows the pseudorapidity distribution of charged hadrons in several centrality classes. We have tuned the normalization of the gluon energy $K_g$ by matching it to the experimental data in the Pb-Pb 0-5\% midrapidity bin~\cite{ALICE:2022imr,ALICE:2010mty}. The {\Dipper} hybrid shows remarkable agreement with data for most of the rapidity range. At large rapidities ($|\eta| > 3$), better agreement is found when including subnuclear flucuations, confirming the interpretation that for smaller (less dense) systems, the subnuclear scales are relevant degrees of freedom.

\noindent \textbf{Net proton multiplicity}: 
In Fig.~\ref{fig:mult}, we show the net proton multiplicity distribution along the rapidity for representative centralities. Including subnuclear fluctuations, reduces the total baryon charge deposited, but does not affect the stopping power (position of the peak). This can be understood from the increased granularity in the initial state, while the peak depends mostly on the energy of the collisions (see Ref. \cite{Garcia-Montero:2024jev}).

\begin{figure}[t]
	\centering
	\includegraphics[width=0.8\textwidth]{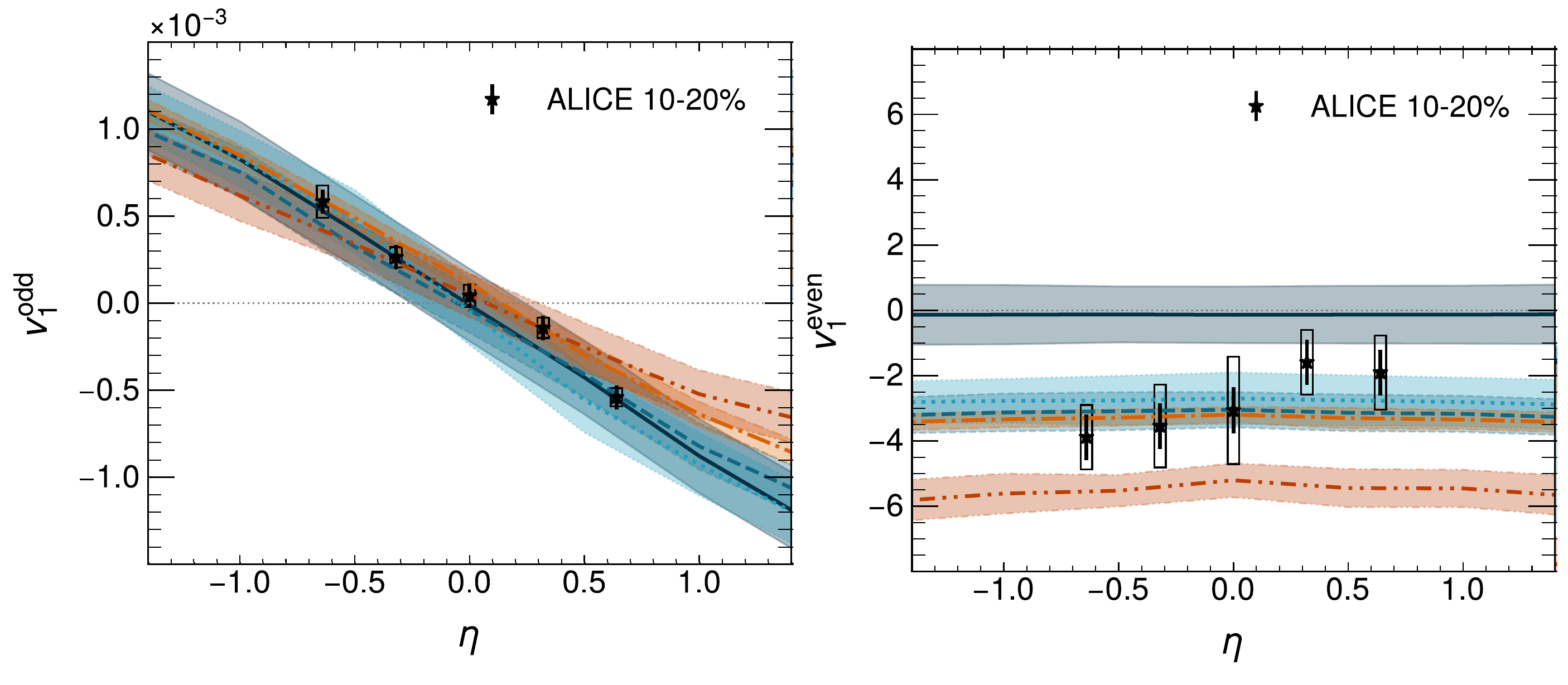}
	\caption{Odd (\textit{left}) and even (\textit{right}) components of directed flow $v_1(\eta)$ of charged hadrons compared with ALICE data \cite{ALICE:2013xri} for the 10-20\% centrality class. }
	\label{fig:v1}\vspace{-8mm}
\end{figure}

\noindent \textbf{Directed flow, $v_1$}: 
In this work, the $v_1$ is calculated relative to the spectator plane as in Ref.~\cite{ALICE:2013xri}. 
Directed flow $v_1$, shown in Fig.~\ref{fig:v1}, is separated into odd and even components relative to the spectator plane. The odd 
component, reflecting the tilted collision geometry, is reproduced well in all cases. The even component, which originates from 
event-by-event asymmetries, is underestimated for scenario (i) but the inclusion of  fluctuations significantly enhanced it. 

\begin{figure}[t]
	\centering
	\includegraphics[width=\textwidth]{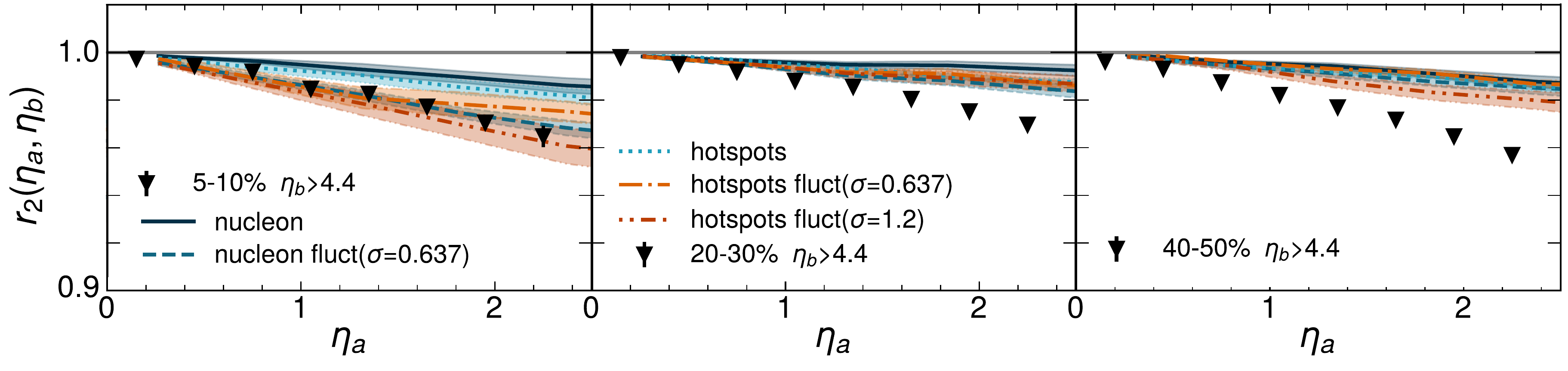}
	\caption{Directed flow $v_1(\eta)$ of charged hadrons compared with ALICE data~\cite{ALICE:2016tlx}. Subnucleonic hotspots enhance the even component, 
		improving agreement with experiment.}
	\label{fig:flow}\vspace{-8mm}
\end{figure}

\noindent \textbf{Flow decorrelation}:
In Fig.~\ref{fig:flow} quantified elliptic flow decorrelations using the standard forward–backward ratio $r_n(\eta_a,\eta_b)$~\cite{CMS:2015xmx}
 the reference rapidity is set to $4.4<\eta_b<5$. Scenario (i) underestimates $r_n$ observed in CMS data, particularly for midcentral and peripheral collisions where geometry-driven flow dominates. Including subnucleonic fluctuations increases decorrelations, especially in central event.Our framework still falls short of experimental results in peripheral collisions, which we attribute to missing sources of fluctuations in the quark sector, which are most relevant in the fragmentation regions.\vspace{-2mm}

\section{Conclusions and Outlook}
In this short communication, we presented first results from the MCDIPPER+CLVisc framework including subnucleonic fluctuations. We focused on a representative set of observables, namely charged particle and  net-proton multiplicities, directed flow and flow decorrelation. Generally, we found that the framework provides rather good descriptions of these observables  without any tuning of transport coefficients for hydrodynamics evolution. Additionally, we have noted that subnucleonic fluctuations have evident effects on the $\eta$-dependence of final state observables, which is clear in all the observables presented.  These findings highlight the relevance of nucleon substructure in shaping the longitudinal dynamics of heavy-ion collisions. Future work 
will incorporate valence-quark fluctuations, a realistic pre-equilibrium stage, and hadronic afterburners for quantitative precision. 
\vspace{-2mm}

\section*{Acknowledgements}
 This work is supported by the Deutsche Forschungsgemeinschaft (DFG, German Research Foundation) through the CRC-TR 211 ``Strong-interaction matter under extreme conditions'' Project No. 315477589--TRR 211. O.G.M. and S.S. also acknowledge support by the German Bundesministerium für Bildung und Forschung (BMBF) through Grant No. 05P21PBCAA. J.Z. is also supported in part by China Scholarship Council (CSC) under Grant No. 202306770009. 
Numerical simulations presented in this work were performed at the Paderborn Center for Parallel Computing ($\rm PC^2$).

%
%


\begin{thebibliography}{99}

\bibitem{Jacak:2012dx}
B.~V.~Jacak and B.~Muller,
Science \textbf{337} (2012), 310-314

\bibitem{Elfner:2022iae}
H.~Elfner and B.~M{\"u}ller,
J. Phys. G \textbf{50} (2023) no.10, 103001

\bibitem{Bozek:2010vz}
P.~Bozek, W.~Broniowski and J.~Moreira,
Phys. Rev. C \textbf{83} (2011), 034911

\bibitem{ALICE:2010mty}
K.~Aamodt \textit{et al.} [ALICE],
Eur. Phys. J. C \textbf{68} (2010), 345-354

\bibitem{PHENIX:2018hho}
A.~Adare \textit{et al.} [PHENIX],
Phys. Rev. Lett. \textbf{121} (2018) no.22, 222301

\bibitem{Garcia-Montero:2025hys}
O.~Garcia-Montero and S.~Schlichting,
Eur. Phys. J. A \textbf{61} (2025) no.3, 54

\bibitem{Garcia-Montero:2023gex}
O.~Garcia-Montero, H.~Elfner and S.~Schlichting,
Phys. Rev. C \textbf{109} (2024) no.4, 044916

\bibitem{Garcia-Montero:2023opu}
O.~Garcia-Montero, H.~Elfner and S.~Schlichting,
PoS \textbf{HardProbes2023} (2024), 054
\bibitem{Garcia-Montero:2025bpn}
O. Garcia-Montero, S. Schlichting, and J. Zhu, Phys. Rev. D \textbf{111}, 076029 (2025).


\bibitem{Wu:2021fjf}
X.~Y.~Wu, G.~Y.~Qin, L.~G.~Pang and X.~N.~Wang,
Phys. Rev. C \textbf{105} (2022) no.3, 034909
\bibitem{Pang:2018zzo}
L.~G.~Pang, H.~Petersen and X.~N.~Wang,
Phys. Rev. C \textbf{97} (2018) no.6, 064918
\bibitem{ALICE:2015bpk}
J.~Adam \textit{et al.} [ALICE],
Phys. Lett. B \textbf{754} (2016), 373-385

\bibitem{ALICE:2013xri}
B.~Abelev \textit{et al.} [ALICE],
Phys. Rev. Lett. \textbf{111} (2013) no.23, 232302

\bibitem{Garcia-Montero:2024jev}
O.~Garcia-Montero and S.~Schlichting,
Phys. Rev. C \textbf{111} (2025) no.2, 024912

\bibitem{ALICE:2016tlx}
J.~Adam \textit{et al.} [ALICE],
Phys. Lett. B \textbf{762} (2016), 376-388

\bibitem{ALICE:2022imr}
S.~Acharya \textit{et al.} [ALICE],
Phys. Lett. B \textbf{845} (2023), 137730
\bibitem{CMS:2015xmx}
V.~Khachatryan \textit{et al.} [CMS],
Phys. Rev. C \textbf{92} (2015) no.3, 034911
\end{thebibliography}
\end{document}